\documentclass[12pt,twocolumn]{iopart}

\usepackage{graphicx}
\usepackage{epsfig}
\usepackage{amssymb}

\begin{document}

\title{Slow down of a globally neutral relativistic $e^-e^+$ beam shearing the vacuum}

\author{E P Alves$^1$, T Grismayer$^1$, M. G. Silveirinha$^{2}$, R A Fonseca$^{1,3}$ and L O Silva$^1$}
\address{$^1$ GoLP/Instituto de Plasmas e Fus\~ao Nuclear - Laborat\'orio Associado, Instituto Superior T\'ecnico, Lisbon, Portugal}
\address{$^2$ Universidade de Coimbra, Departmento de Engenharia Electrot\'ecnica, Instituto de Telecomunica\c{c}\~oes, 3030-290 Coimbra, Portugal}
\address{$^3$ DCTI/ISCTE Instituto Universit\'{a}rio de Lisboa, 1649-026 Lisboa, Portugal}
\eads{\mailto{e.paulo.alves@ist.utl.pt}}
\eads{\mailto{luis.silva@ist.utl.pt}}

\date{\today}

\begin{abstract}
The microphysics of relativistic collisionless sheared flows is investigated in a configuration consisting of a globally neutral, relativistic $e^-e^+$ beam streaming through a hollow plasma/dielectric channel. We show through multidimensional PIC simulations that this scenario excites the Mushroom instability (MI), a transverse shear instability on the electron-scale, when there is no overlap (no contact) between the $e^-e^+$ beam and the walls of the hollow plasma channel. The onset of the MI leads to the conversion of the beam's kinetic energy into magnetic (and electric) field energy, effectively slowing down a globally neutral body in the absence of contact. The collisionless shear physics explored in this configuration may operate in astrophysical environments, particularly in highly relativistic and supersonic settings where macroscopic shear processes are stable.
\end{abstract}

\maketitle

\section{Introduction}

Sheared flow configurations are pervasive in nature. Their study is of fundamental importance since they trigger instabilities and drive turbulence both in ionized and non-ionized settings. While the stability and evolution of sheared flow configurations has been thoroughly studied at the macroscopic (hydrodynamic/magnetohydrodynamic) level \cite{1961hhs..book.....C,Frank:1996wn,Keppens:1999vu,Bodo:2004id,Zhang:2009kb,Hamlin:2013ca}, their dynamics at the microscopic (plasma kinetic) level remain poorly understood. Recent theoretical and particle-in-cell (PIC) simulation studies have revealed that unmagnetized sheared flow configurations host a wealth of collisionless electron-scale instabilities \cite{Alves:2012vb,Grismayer:2013bi,Bussmann:2013dt,Liang:2013gv,Alves:2014dm,Nishikawa:2014ct,Alves:2015wo}. These include the electron-scale Kelvin-Helmholtz instability (ESKHI), which develops parallel to the flow \cite{Alves:2012vb,Grismayer:2013bi,Bussmann:2013dt,Liang:2013gv,Alves:2014dm,Nishikawa:2014ct}, and the Mushroom instability (MI), which develops in the transverse direction to the flow \cite{Alves:2015wo}. Both the ESKHI and MI result from the coupling between high-frequency electromagnetic waves and electron plasma waves in the presence of a velocity shear. These microscopic instabilities can play an important role in ultra-relativistic and highly supersonic sheared flow scenarios, where the (magneto)hydrodynamic Kelvin-Helmholtz instability (KHI) is found to be stable \cite{Bodo:2004id,Hamlin:2013ca}.

Moreover, due to their electromagnetic character, both the ESKHI and MI can operate in the presence of a finite vacuum gap between shearing flows, i.e., in the absence of "contact" \cite{Alves:2015wo}. This effect is closely related to the development of electromagnetic instabilities in shearing, globally neutral, polarizable metallic/dielectric slabs separated by a nanometer-scale gap, which results in an effective non-contact friction force between the slabs \cite{Silveirinha:2014jh,Silveirinha:2014ec}. This form of non-contact friction, associated with the development of electromagnetic instabilities, is the classical analogue to the quantum friction effect proposed by Pendry \cite{1997JPCM....910301P}. While the effect of non-contact friction has been considered in the sub-relativistic regime, connected to the development of electromagnetic modes parallel to the flow (ESKHI), the role of electromagnetic modes transverse to the flow has been overlooked. In this work, we show that the transverse modes are dominant in relativistic settings, and are responsible for the non-contact friction force in the relativistic regime.

Reproducing in the laboratory the collisionless conditions that characterize most astrophysical systems is highly challenging. Sheared flow experiments have been performed in the collisional regime \cite{Hurricane:2009km,Harding:2009fc}, revealing the development of the hydrodynamic KHI under high-energy-density conditions. More recently, the experimental observation of sheared flow driven turbulence in the collisionless regime was reported \cite{Kuramitsu:2012fs}, but clear signatures of the ESKHI/MI were not observed. We believe that the ESKHI/MI processes may be better observed in configurations where bulk overlap between the flows is minimized, inhibiting streaming instabilities that could mask their effects. Relativistic, low divergence and globally neutral plasma flows may be achieved by superimposing relativistic electron ($e^-$) and positron ($e^+$) bunches from linear accelerators \cite{Muggli:2013wh}. Alternatively, the generation of energetic, globally neutral $e^-e^+$ beams has also been achieved resulting from QED pair cascades triggered by the collision of a laser-wakefield-accelerated electron bunch with a solid target \cite{Sarri:2015jq}. The production of such exotic beams will enable the laboratory exploration of collisionless microphysical processes in the relativistic regime, which are directly relevant to many astrophysical systems. In particular, it has been suggested that the interaction of such beams with a bulk stationary plasma would allow to probe the collisionless current filamentation instability (CFI) in the relativistic regime \cite{Muggli:2013wh,Sarri:2015jq}. Here, we investigate a configuration where an $e^-e^+$ beam streams through a stationary hollow plasma channel, or hollow dielectric, in order to probe the microphsyics of relativistic collisionless sheared flows.

In this work, we investigate the relativistic collisionless shear dynamics associated with a relativistic, globally neutral $e^-e^+$ beam streaming through a hollow plasma channel. The $e^-e^+$ beam is considered to be propagating with a relativistic Lorentz factor $\gamma_0$, and to have a gaussian density profile with $\sigma_\perp = 2\sigma_\parallel=2c/\omega_{pe}$ ($\perp$ and $\parallel$ are perpendicular and parallel to the propagation direction, respectively), and a peak number density $n_b$ (measured in the laboratory frame); $c$ denotes the speed of light in vacuum, and $\omega_{pe}=\sqrt{e^2n_b/m_e\epsilon_0}$ where $\sqrt{2}\omega_{pe}$ is the plasma frequency of the beam (the $\sqrt{2}$ factor arises from the positron contribution). These beam parameters have been considered in the context of exploring the CFI in the relativistic regime \cite{Muggli:2013wh}, and can be made available at SLAC National Accelerator Laboratory with $\gamma_0 \sim \mathcal{O}(10^4)$. The hollow plasma channel has an internal radius $R$ where the plasma density is assumed to be 0, and the surrounding plasma density is $n_{p}=n_b$. This channel could also be made of a hollow solid-state dielectric. The impact of the dielectric properties of the hollow channel on the development of the shear dynamics will be explored elsewhere. In this work, we assume the channel walls to have the dielectric response of a cold collisionless plasma with the same density as the beam.

We first explore the changes in the dynamics of the propagation of the relativistic $e^-e^+$ beam in channels with different radii $R$. In particular, we aim to clearly distinguish the dynamics in narrow hollow channels ($R/\sigma_\perp \sim 1$), where significant overlap between the beam and the walls of the hollow channel occurs, triggering bulk streaming instabilities like the current filamentation instability (CFI) \cite{Muggli:2013wh,Sarri:2015jq}, from the dynamics in wide hollow channels ($R/\sigma_\perp \gtrsim 3$), where bulk overlap is negligible and only collisionless sheared flow instabilities may develop. We thus begin by reviewing the growth rates of the relevant instabilities in order to understand their relative importance under different conditions.

\section{Bulk streaming instabilities versus shear instabilities}

The the development of the CFI triggered by a relativistic $e^-e^+$ beam propagating in a stationary bulk plasma has been explored by \cite{Muggli:2013wh,Sarri:2015jq}. The CFI growth rate in the relativistic limit scales as $\Gamma_\mathrm{CFI}/\omega_{pe} \sim \gamma_0^{-1/2}$. It is expected that if significant overlap between the beam and the walls of plasma channel exists, the CFI will develop with $\Gamma_\mathrm{CFI}$ calculated using the average overlapping beam density.

We have shown elsewhere \cite{Alves:2012vb,Alves:2014dm,Alves:2015wo} that unmagnetized collisionless sheared flows can trigger both the ESKHI (along the sheared flow) and the MI (transverse to the sheared flow). Although the finite geometry of the beam will certainly be important, we can estimate the growth rates of both MI and ESKHI modes based on a slab geometry for the sheared flow. Assuming a semi-infinite relativistic plasma slab propagating with $\gamma_0$, shearing with a second semi-infinite stationary plasma slab, the respective growth rates scale as $\Gamma_\mathrm{ESKHI}/\omega_{pe} \sim \gamma_0^{-3/2}$ \cite{Alves:2012vb,Alves:2014dm} and $\Gamma_\mathrm{MI}/\omega_{pe} \sim \gamma_0^{-1/2}$ \cite{Alves:2015wo}. It is therefore expected that under highly relativistic conditions, relevant to the $e^-e^+$ beam streaming in a hollow channel scenario, the transverse MI will be dominant due to its weaker dependence on $\gamma_0$. In addition, the length of the $e^-e^+$ beam ($\sim \sigma_\parallel = c/\omega_{pe}$) is much shorter than the wavelength of the most unstable mode of the ESKHI ($\sim \bar{\gamma_0}^{3/2}c/\omega_{pe}$), so the ESKHI should not develop at all. Furthermore, due to their electromagnetic nature, these microscopic shear instabilities can operate in the absence of "contact", i.e. when there is a finite gap of vacuum between the sheared flows \cite{Alves:2015wo}, which can be directly explored in the hollow plasma channel configuration. Using the appropriate boundary conditions to describe the shear interaction between an ultra relativistic and stationary plasma slabs, separated by a finite gap $L_g$, one finds that the growth rate of the MI scales as $1/L_g$.

\section{PIC simulations of a relativistic $e^-e^+$ beam streaming through a hollow plasma channel}

We analyze the effect of different diameters of the hollow plasma channel on the dynamics of relativistic $e^-e^+$ beam. We have performed 2D PIC simulations of the transverse plane to the propagation direction using the PIC code OSIRIS \cite{Fonseca:2002wg,Fonseca:2008ib}. Although the 2D simulations assume both an infinitely long beam and hollow plasma channel, these simulations capture the transverse electron-scale shear dynamics and the interplay with the CFI in the case of overlap. We simulate a domain with dimensions $L_x\times L_y=26^2~(c/\omega_{pe})^2$, resolved with $256^2$ cells; $36$ particles per cell per species were used. The $e^-e^+$ beam is positioned at the centre of the box and propagates out of the plane ($+z$ direction). The walls of the cylindrical plasma channel are placed at a distance $R$ from the centre of the beam. The plasma composing the walls of the hollow plasma channel has the same electron density as the beam, and contains heavy ions (assumed immobile in the simulations).

\subsection{Strong overlap}
\begin{figure}[t!]
\begin{center}
$$\includegraphics[width=0.5\columnwidth]{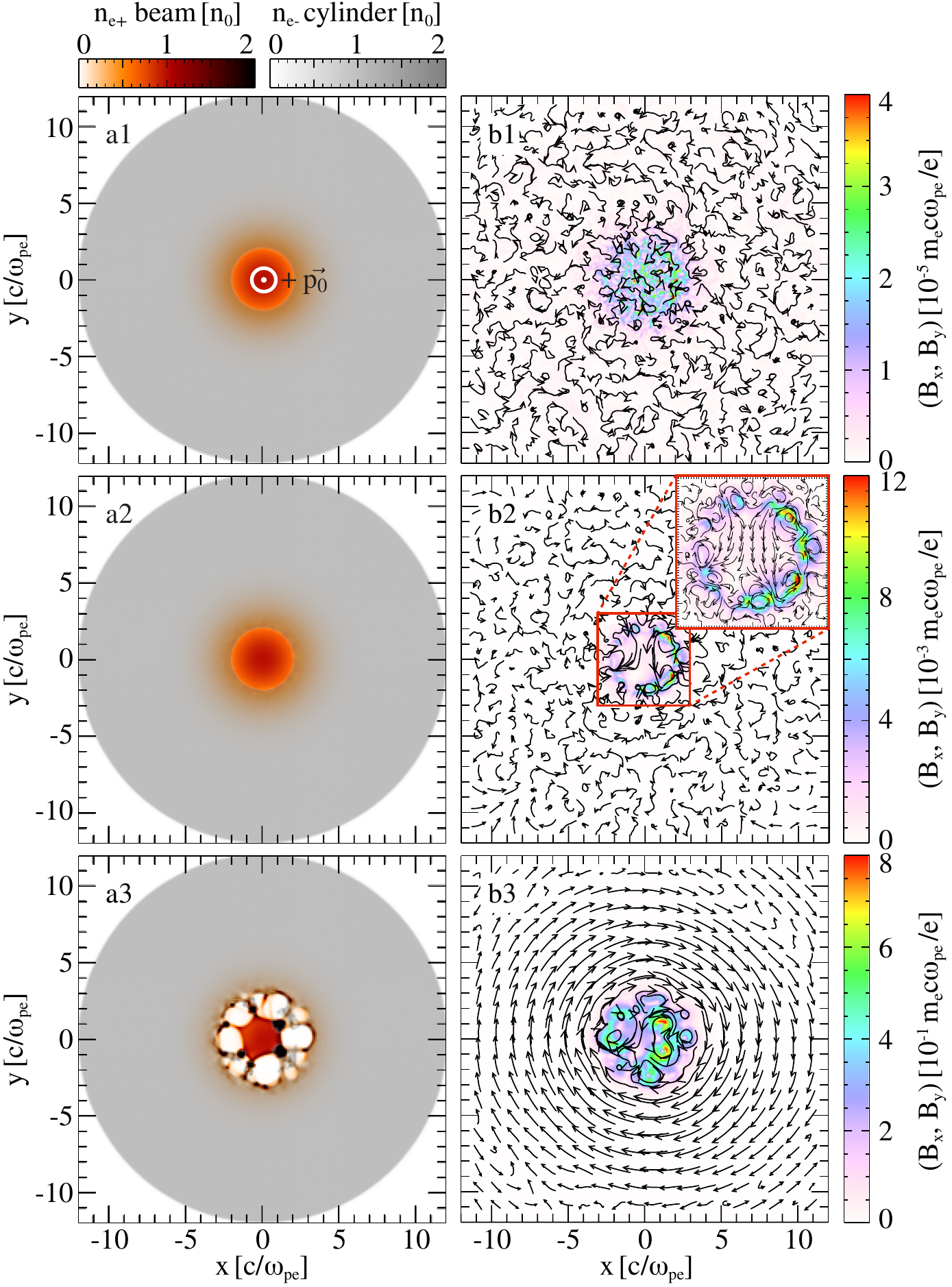}$$ 
\caption{
Evolution of (a) number density and (b) self-generated magnetic field in the interaction between a relativistic $e^-e^+$ beam ($\gamma_0=100$, propagating out of the plane) and a narrow hollow plasma channel (at rest) with $R=\sigma_\perp$, illustrating the development of the bulk CFI (inset of b2) due to the strong overlap between the beam and the plasma channel. Red and grey scales correspond to the positron number density of the beam and the electron density of the hollow plasma channel, respectively. Evolution of the interaction is shown at (1) $\omega_{pe}t = 20$ (initial setting), (2) $\omega_{pe}t = 80$ (linear regime) and (3) $\omega_{pe}t = 140$ (nonlinear regime/beam breakup).
}
\label{fig:1}
\end{center}
\end{figure}

Figure~\ref{fig:1} illustrates the case of strong overlap between an $e^-e^+$ beam (with $\gamma_0=10^2$) and the walls of a narrow hollow plasma channel with $R=\sigma_\perp=2c/\omega_{pe}$. The evolution of the positron density of the $e^-e^+$ beam and the electron density of the walls of the plasma channel is shown in Figure~\ref{fig:1}~(a), and the evolution of the self-generated magnetic field is shown in Figure~\ref{fig:1}~(b). The strong overlap between the beam and the walls of the channel is clearly observed in Figure~\ref{fig:1}~(a1). The initial condition of current neutrality is verified in the magnetic field structure of Figure~\ref{fig:1}~(b1), which shows small amplitude and small-scale magnetic fluctuations due to thermal noise of the beam. At $\omega_{pe}t=80$ the development of magnetic field structures associated with the CFI becomes clearly visible (inset of Figure~\ref{fig:1}~(b2)); these structures occur precisely in the region of overlap between the beam and the walls of the plasma channel, creating the bulk counter-streaming conditions required to trigger the CFI \cite{Muggli:2013wh}. The growth rate of the CFI, measured from the evolution of the total magnetic field energy in the simulation domain, matches the theoretical growth rate when taking into account the density of the beam in the overlapping region. The nonlinear development of the CFI ultimately leads to the break-up of the $e^-e^+$ beam into multiple current filaments as shown in Figures~\ref{fig:1}~(a3) and (b3).

\subsection{Weak overlap}
\begin{figure}[t!]
\begin{center}
$$\includegraphics[width=0.5\columnwidth]{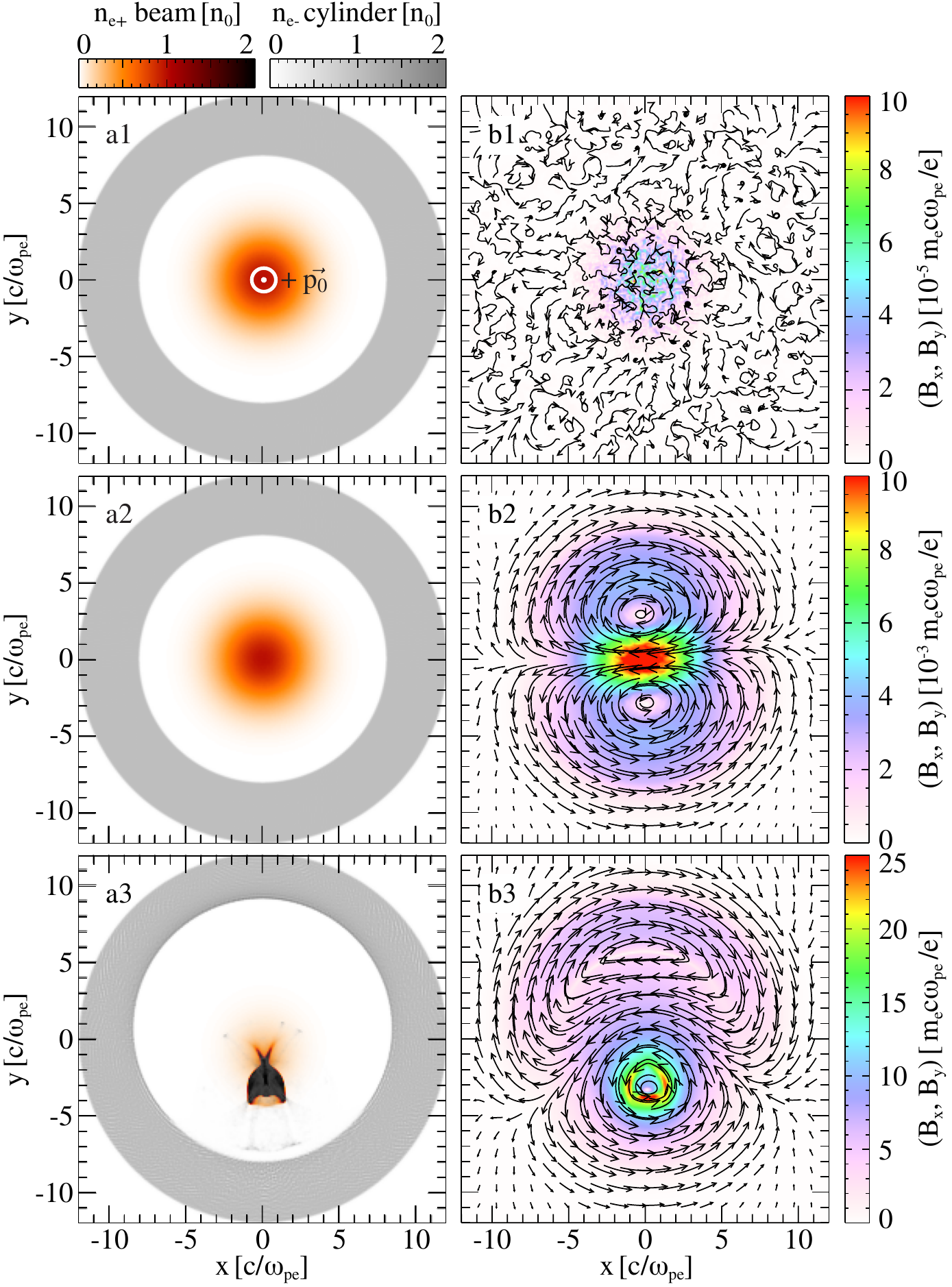}$$ 
\caption{
Evolution of (a) number density and (b) self-generated magnetic field in the interaction between a relativistic $e^-e^+$ beam ($\gamma_0=10^2$, propagating out of the plane) and a wide hollow plasma channel (at rest) with $R=4\sigma_\perp$ (negligible overlap between beam and plasma channel), illustrating the development of the transverse relativistic shear instability (MI). Evolution of the interaction is shown at (1) $\omega_{pe}t = 50$ (initial setting), (2) $\omega_{pe}t = 600$ (linear regime) and (3) $\omega_{pe}t = 850$ (nonlinear regime/beam breakup).
}
\label{fig:2}
\end{center}
\end{figure}

For channels with $R\gtrsim4\sigma_\perp$, the overlapping beam density with the walls of the plasma channel becomes very weak, inhibiting the development of the CFI. A case of weak overlap is illustrated in Figure~\ref{fig:2} where the channel radius has $R/\sigma_\perp=4$. All other parameters are kept fixed relative to the case of Figure~\ref{fig:1}. Again, the system is initially current and charge neutral, and no background fields are present. Only small electromagnetic fluctuations are present in the early stage of the simulation, associated with the finite emittance of the relativistic beam. At $\omega_{pe}t=600$, a self-generated magnetic dipole structure emerges as shown in Figure~\ref{fig:2}~(b2). This is a signature of the transverse electron-scale shear instability, the MI. For small $\sigma_\perp$, as used in these simulations, the dipole mode is found to be the fastest growing mode. Higher order modes have been observed in simulations with larger values of $\sigma_\perp$, while fixing $R/\sigma_\perp=4$. A predictive analytic theory describing this configuration is still lacking, and will be developed in a future publication. The continued growth of the dipole magnetic field structure due to the MI, eventually leads to the breakup of the $e^-e^+$ beam (Figures \ref{fig:2}~(a3) and (b3)). If the beam escapes the channel before breakup, the dipolar magnetic field structure can be preserved and sustained by the beam on a time-scale $~1/(\omega_{pe}/\sqrt{\gamma_0})$. This configuration can be used to imprint a tailored field structure in a globally neutral relativistic beam.

\begin{figure}[t!]
\begin{center}
$$\includegraphics[width=1.0\columnwidth]{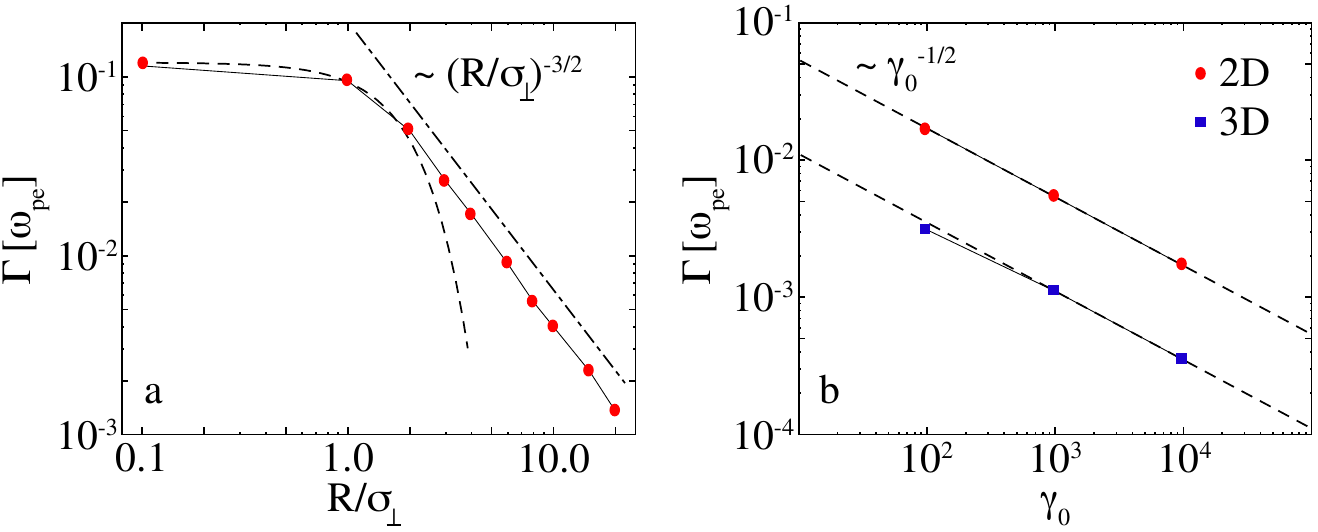}$$ 
\caption{
\textit{Left}: growth rate dependence of the self-generated magnetic field on the inner radius $R$ of the hollow plasma channel, showing transition from CFI to MI. Here, the $e^-e^+$ beam has $\gamma_0=10^2$ and a fixed transverse width of $\sigma_\perp=2c/\omega_{pe}$. Red points represent results of 2D simulations. Dashed curve represents theoretical growth rate of the CFI considering the beam density at overlap. \textit{Right}: Growth rate of the MI-generated magnetic field versus $e^-e^+$ beam Lorentz factor $\gamma_0$. Here, the inner radius of the hollow channel is set to $R=4\sigma_\perp$ (weak overlap/no CFI), with $\sigma_\perp=2c/\omega_{pe}$. Red and blue points represent results of 2D and 3D PIC simulations respectively.
}
\label{fig:3}
\end{center}
\end{figure}

The growth rate of the total self-generated magnetic field energy (measured from 2D simulations) for different hollow channel radii ($R$) is shown in Figure~\ref{fig:3} (a). The cases with $R/\sigma_\perp < 3$ are characterized by strong overlap between the beam and the walls of the hollow channel. The growth rates are found to agree with the theoretical prediction for the CFI (dashed line), taking into account the beam density in the overlapping region. At $R/\sigma_\perp = 3$, magnetic field generation activity occurs both in the overlapping region, due to the CFI, and at the beam position, due to the MI, marking the transition to the collisionless shear-dominated regime. For $R/\sigma_\perp \ge 4$, the measured growth rates are higher than predicted by the CFI curve, which decays abruptly due to the low beam density at overlap. Here, the dynamics of the system is dominated by the collisionless shear dynamics of the MI, whose growth rate is found to decay as $(R/\sigma_\perp)^{-3/2}$.

Figure~\ref{fig:3} (b) shows the dependence of the MI growth rate on the relativistic Lorentz factor ($\gamma_0$), for a channel with $R/\sigma_\perp = 4$, taken from 2D and 3D PIC simulations. The growth rates measured from 3D simulations (blue points) are $\approx 5$ times lower than those measured from 2D simulations (red points). This numerical factor arises from the spatio-temporal development of the MI in the longitudinally finite ($\sigma_\parallel = 1c/\omega_{pe}$) beam in 3D. Nevertheless, the MI growth rate is found to decay in both cases with $\gamma_0^{-1/2}$, which is consistent with the theoretical model for the slab geometry \cite{Alves:2015wo}.

\begin{figure}[t!]
\begin{center}
$$\includegraphics[width=1.0\columnwidth]{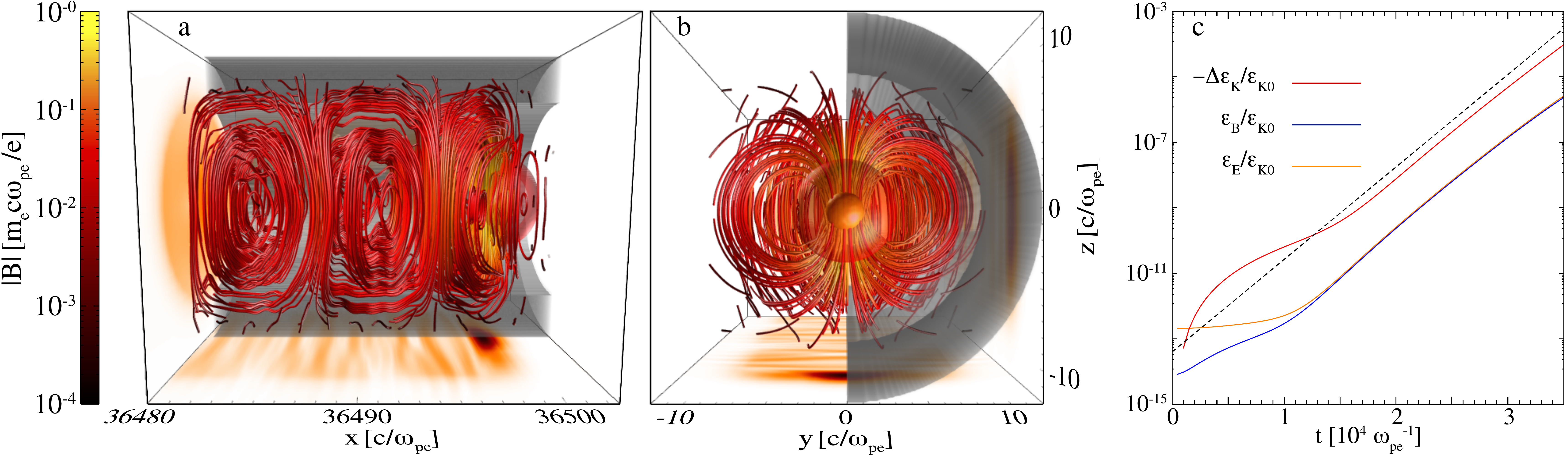}$$ 
\caption{
3D PIC simulation of the development of the MI for $\gamma_0=10^4$, $\sigma_\perp=2\sigma_\parallel=2c/\omega_{pe}$ and $R=4\sigma_\perp$. The $e^-e^+$ beam density is represented by the red isosurfaces, and the hollow plasma channel is represented by the grey isosurfaces (half of the cylindrical structure has been omitted to more clearly illustrate the dynamics of the $e^-e^+$ beam inside). The $e^-e^+$ beam propagates in the positive $x$ direction. The red lines represent the structure of the MI-generated magnetic field. Figures (a) and (b) are side and front views of the simulation domain, respectively, taken at $\omega_{pe}t = 36500$. Figure (c) shows the temporal evolution of the electric (orange curve) and magnetic (blue curve) field energy and the negative variation of the total kinetic energy of the beam (red curve), all normalized to the initial kinetic energy of the beam. The slope of the dashed line represents the growth rate of the MI.
}
\label{fig:4}
\end{center}
\end{figure}

The fully self-consistent structure of the MI-generated magnetic fields from a 3D PIC simulation is illustrated in Figure~\ref{fig:4}. The simulation domain has dimensions $L_x\times L_y\times L_z=26^3(c/\omega_{pe})^3$, resolved with $256^3$ cells. The red/orange isosurfaces represent the positron density of an ultra relativistic $e^-e^+$ beam with $\gamma_0 = 10^4$, propagating in the $+x$-direction inside a hollow plasma channel (grey isosurfaces) with $R/\sigma_\perp = 4$. Half of the plasma channel wall has been omitted to observe the growing magnetic field structure inside. The MI-generated dipolar magnetic field structure, previously identified in 2D simulations, is clearly observed at the $e^- e^+$ beam position in the 3D simulation (Figure~\ref{fig:4} (b)); the field lines are bowed towards the back of the beam in 3D (Figure~\ref{fig:4} (a)). The growing magnetic field is sustained by a current structure associated with the charge separation between electrons and positrons of the beam. This charge separation also induces a dipolar electric field structure, which triggers the trailing wake-field structure behind the beam (Figure~\ref{fig:4} (a)). For an $e^-e^+$ beam density of $n_b=10^{17}~\mathrm{cm}^{-3}$, the snap shots of Figure~\ref{fig:4} correspond to a propagation length of $\sim 50~\mathrm{cm}$.

The generation of electric and magnetic field energy due to the MI is accompanied by the self-consistent loss in kinetic energy of the beam. This is illustrated in Figure~\ref{fig:4} (c), which shows the exponential growth of electric and magnetic field energy, and the corresponding negative variation of the beam's kinetic energy. The development of the MI therefore acts as an effective non-contact friction force, which is proportional to the MI growth rate, $d p_\parallel/dt = - 2 \Gamma_\mathrm{MI}~p_\parallel = D_\mathrm{n.c.}$ ($p_\parallel$ is the average longitudinal momentum of the beam). The effective non-contact friction force ($D_\mathrm{n.c.}$) associated with the development of electromagnetic instabilities between shearing dielectrics separated by a vacuum gap has been previously explored in \cite{Silveirinha:2014jh,Silveirinha:2014ec}, where the longitudinal electromagnetic modes played the most important role at subrelativistic velocity shears. Here, we have shown that the transverse MI modes are responsible for the effective non-contact friction effect in the relativistic regime.

\subsection{Non ideal $e^-e^+$ beam and off-axis propagation}

If the $e^-e^+$ beam is not perfectly charge neutral ($n_{e^+}-n_{e^-} = \delta n \ne 0$) it will defocus on a time-scale $\omega_{pe}\tau_\mathrm{defocus} \sim \gamma_0^{3/2}(n_b/\delta n)^{1/2}$, where $n_b$ is $\mathrm{min}(n_{e^+},n_{e^-})$. Fortunately, for ultra relativistic beams, this defocusing time is much larger than the growth time of the MI $\omega_{pe}\tau_\mathrm{MI} \sim \omega_{pe}/\Gamma_\mathrm{MI} \propto \gamma_0^{1/2}$, allowing for the development of the MI before the beam defocuses. We have observed the persistent development of the MI in beams with $\delta n=0.1n_b$ in 3D PIC simulations, where the MI-generated fields were found to be superposed over the self-consistent fields associated with the global current of the non-neutral beam.

We have also simulated the effect of the $e^-e^+$ beam propagating off-axis, at a finite distance $r_0$ from the axis of the hollow channel. We find for offsets of $r_0/R \leq 1/4$, while still preserving the weak overlap condition $(R-r_0)/\sigma_\perp \ge 4$, the continual development of the MI with only small variations in the growth rate $<10\%$ relative to the ideal on-axis case. The emerging dipolar field structure becomes slightly asymmetric under these conditions.

\section{Conclusion}
In conclusion, we have shown that the transverse microphysics of relativistic collisionless sheared flows can be investigated in a scenario where a relativistic, globally neutral $e^-e^+$ beam propagates inside a (stationary) hollow plasma channel. Resorting to 2D PIC simulations, we have studied the propagation of a relativistic $e^-e^+$ beam in different size hollow plasma channels. We have found that for channel radii with $R/\sigma_\perp \lesssim 3$ strong overlap between the channel and beam occurs, exciting the CFI in the overlapping region, whereas for channel radii with $R/\sigma_\perp \gtrsim 4$ the overlap is weak, inhibiting the CFI and allowing for the clear development of the relativistic collisionless shear dynamics of the MI. For the range of $e^-e^+$ beam and hollow channel parameters explored in this work, the MI growth rate was found to scale as $\Gamma_\mathrm{MI}/\omega_{pe} \propto (\gamma_0R^3/\sigma_\perp^3)^{-1/2}$. Interestingly, despite the microscopic nature of this shear instability, we find the persistent onset of the MI even when the gap between the beam and the walls of the plasma channel are $R/\sigma_\perp \gg 1$. These scalings also support the development of the MI when the $e^-e^+$ beam is not perfectly neutral, since the MI growth time is shorter than the electrostatic defocusing time. The full spatio-temporal evolution of the MI was observed via 3D PIC simulations, having revealed the self-consistent structure of the MI-generated magnetic field. The self-consistent slow down of the relativistic $e^-e^+$ beam, due to the conversion of the beam's kinetic energy into electric/magnetic field energy through the MI was also verified. The development of the MI therefore leads to an effective non-contact friction force which is proportional to $\Gamma_\mathrm{MI}$. These are the first \textit{ab initio} 3D PIC simulations capturing the self-consistent slow down of a globally neutral body in the absence of contact \cite{Silveirinha:2014jh,Silveirinha:2014ec,1997JPCM....910301P}.

\ack

The authors would like to thank J. Vieira for fruitful discussions. This work was partially supported by the European Research Council ($\mathrm{ERC-2010-AdG}$ Grant 267841) and FCT (Portugal) grants SFRH/BD/75558/2010, SFRH/BPD/75462/2010, and PTDC/FIS/111720/2009. We acknowledge PRACE for awarding access to SuperMUC based in Germany at Leibniz research center. Simulations were performed on the ACCELERATES cluster (Lisbon, Portugal), and JUQUEEN (Germany).

\section*{References}

\bibliographystyle{iopart-num.bst}
\bibliography{23-06-2015.bib}


\end{document}